\documentclass[dvips,12pt]{article}
\usepackage{graphicx}
\newcommand{\doublespacing}{\let\CS=\@currsize\renewcommand{\baselinesstrech}
{2.0}\tiny\CS}

\begin{document}

\textwidth 16cm
\newcommand{\bd}{\begin{document}}
\newcommand{\ed}{\end{document}}
\newcommand{\bc}{\begin{center}}
\newcommand{\ec}{\end{center}}
\newcommand{\vs}{\vspace}

\bc {\huge \bf Pseudo supersymmetric Partners } \ec

\bc {\huge \bf for } \ec

\bc {\huge \bf the Generalized Swanson Model } \ec

\vs{.5cm}

\bc
{\it \large A. Sinha{\footnote {e-mail : anjana$_-$t@isical.ac.in}}} \\
\ec

\bc {and} \ec

\bc
{\it \large P. Roy{\footnote {e-mail : pinaki@isical.ac.in}}} \\
\ec

\bc
{\it \large Physics \& Applied Mathematics Unit \\
Indian Statistical Institute \\
Kolkata - 700 108 \\ INDIA } \ec

\vs{.5cm}

\begin{abstract}

New non Hermitian Hamiltonians are generated, as isospectral
partners of the generalized Swanson model, viz., $ H_- =
{\cal{A}}^{\dagger} {\cal{A}} + \alpha {\cal{A}} ^2 + \beta
{\cal{A}}^{\dagger \ 2} $, where $ \alpha \ , \ \beta $ are real
constants, with $ \ \alpha \neq \beta $, and ${\cal{A}}^{\dagger}$
and ${\cal{A}}$ are generalized creation and annihilation
operators. It is shown that the initial Hamiltonian $H_-$, and its
partner $H_+$, are related by pseudo supersymmetry, and they share
all the eigen energies except for the ground state. This pseudo
supersymmetric extension enlarges the class of non Hermitian
Hamiltonians $H_{\pm}$, related to their respective Hermitian
counterparts $h_{\pm}$, through the same similarity transformation
operator $\rho$ : $ H_{\pm} = \rho ^{-1} \ h_{\pm} \ \rho $. The
formalism is applied to the entire class of shape-invariant
models.

\vspace{1cm}


\vspace{1cm}


\end{abstract}

\pagebreak


\section{Introduction}

Ever since interest in non Hermitian Hamiltonians (with real
energies) was revived about a decade ago by Bender and Boettcher
\cite{bender}, quantum systems described by such non Hermitian
Hamiltonians have been studied widely \cite{nonHerm}. To extend
the class of such systems, new exactly solvable (or quasi-exactly
solvable) non Hermitian Hamiltonians with real, discrete energies,
have been generated using different approaches
--- e.g., supersymmetry \cite{susy}, the related intertwining
operator method \cite{intertwine}, or the Darboux algorithm
\cite{darboux}. In a recent work \cite{jpa-07}, we had found a
class of new non Hermitian models by generalizing the Swanson
Hamiltonian $ H = a^{\dagger} a + \alpha a ^2 + \beta a^{\dagger \
2} $ where $ \alpha \ , \ \beta $ are real constants, with $ \
\alpha \neq \beta $. This model was initially proposed by Swanson
\cite{swanson}, and later on studied by various authors
\cite{geyer}. For the sake of generalization, we had used
generalized creation and annihilation operators,
${\cal{A}}^{\dagger}$ and ${\cal{A}}$, in place of Harmonic
oscillator creation and annihilation operators $a^{\dagger} \ , \
a$, so that $ H_- = {\cal{A}}^{\dagger} {\cal{A}} + \alpha
{\cal{A}} ^2 + \beta {\cal{A}}^{\dagger \ 2} $. The energies of
this class of non Hermitian Hamiltonians were found to be real
when the parameters satisfy the relations $(\alpha + \beta) < 1 $
and $ 4 \alpha \beta < 1 $ \cite{jpa-07, swanson}. In the present
work, we shall generate new non Hermitian Hamiltonians $H_+$, as
isospectral partners of the generalized Swanson Hamiltonian $ H_-
$. It may be recalled that non Hermiticity may be introduced
through a {\it scalar} term or by a {\it vector} term. In the {\it
scalar} case non Hermiticity may be introduced by replacing $x $
with $(x \pm i \epsilon) $ or by taking one of the parameters
complex in the expression for the potential $V(x)$. Though this
looks apparently simple, nevertheless, a similarity transformation
of the non Hermitian Hamiltonian ($H_{\pm}$ with solutions $\psi
^{(\pm)}$) maps it to a complicated non-local Hermitian
Hamiltonian ($h_{\pm}$ with solutions $\phi ^{(\pm)}$).
Consequently, the resulting Hermitian Hamiltonian is not exactly
solvable and perturbative techniques may have to be applied for
solving the same. Additionally, it may not be possible to
determine the exact form of the metric operator $\eta $
explicitly, with respect to which the inner product $< \psi _m |
\eta | \psi _n > $ is positive definite. On the contrary, if non
Hermiticity is introduced through an imaginary {\it vector }
potential, the similarity transformation yields a Hermitian
Schr\"{o}dinger Hamiltonian (consisting of the standard kinetic
term plus a local Hermitian potential term), with possibilities of
exact (or quasi-exact) solvability. The metric operator also can
be obtained in closed form. In view of the fact that a gauge-like
transformation transforms the non Hermitian Hamiltonian $H_-$ to a
Hermitian one $h_-$ with Schr\"{o}dinger form, the imaginary
vector potential may be regarded as a trivial way of introducing
non Hermiticity. Nevertheless, since the resulting Hamiltonian
$H_-$ comes out to be real yet non Hermitian, with real spectrum
and possibilities of exact (or quasi-exact) solutions, our
interest here is to look for isospectral partners of such a
potential.

A couple of recent works deserve special mention here
\cite{quesne1, quesne2}. In the first of these \cite{quesne1}, a
method was proposed to generate a family of non Hermitian
Hamiltonians equivalent to the Swanson Hamiltonian \cite{swanson},
by writing $H$ as a linear combination of $su(1,1)$ generators
$K_0, K_{\pm}$ ; i.e. $ H = 2 K_0 + 2 \alpha K_- + 2 \beta K_+ $.
In the second work \cite{quesne2}, a quasi-Hermitian
supersymmetric extension was proposed for a Harmonic Oscillator
Hamiltonian, augmented by a non Hermitian ${\cal{PT}}$ symmetric
part. To construct new non Hermitian Hamiltonians related by
similarity transformation to Hermitian ones \cite{jones,
mostafa2}, the $su(1,1)$ Lie algebra needs to be enlarged to a
$su(1,1/1) \sim osp(2/2, \mathbf{R}) $ Lie super algebra.
Incidentally, both in ref. \cite{quesne2} and our present
formulation, non Hermiticity is introduced not by considering a
complex-valued potential, but through a momentum-dependent
interaction term. However, our approach is different from either
of those employed in ref. \cite{quesne1, quesne2}. Instead of the
Swanson model considered in ref. \cite{quesne1, quesne2}, we deal
with its generalized version \cite{jpa-07}. So while the non
Hermiticity is introduced through a momentum-dependent {\it
linear} interaction term in ref. \cite{quesne2}, viz., $ i \left(
\alpha - \beta \right) \left( xp + px \right) $; in our case it is
not necessarily linear, to be precise it is of the form $ i \left(
\alpha - \beta \right) \left\{ W(x)p + pW(x) \right\} $, thus
depending on the particular model considered. Secondly, in our
case the Hamiltonian $H_-$ is written in terms of generalized
creation and annihilation operators, which are not necessarily Lie
algebra generators.

It is worth recalling here that the choice of the metric operator
$\eta$ is not unique. In fact many possible such operators exist,
obeying the condition $ \eta H_- = h_- ^{\dagger} \eta $. So there
are many ways of finding a Hermitian Hamiltonian $h_-$, through a
similarity transformation $ h_- = \rho H_- \rho ^{-1}$ and thus
obtain the metric operator $\eta$ \cite{mostafa}, from the
relation $\eta = \rho ^2$ \cite{jones}. Each $h_-$ is associated
with a different metric, thus invoking a different Hilbert space
for each Hermitian map. Our approach gives a simple,
straightforward method to determine one such similarity
transformation $\rho$ mapping the non Hermitian system $H_-$ to
its Hermitian equivalent $ h_- $. In this respect our approach is
different from \cite{quesne2} where the $su(1,1)$ generators $K_0,
K_{\pm}$ are used in the construction of $\rho$. Furthermore, the
normalization requirement of the wave functions for pseudo
Hermitian systems viz., $ < \psi \mid \eta \mid \psi > $ ensures
the wave functions to be naturally normalized in our formalism, as
we shall see later.

Once a non Hermitian partner Hamiltonian $H_+$ of the generalized
Swanson Hamiltonian $H_-$ is obtained, it is natural to look for
some underlying symmetry between $H_{\pm}$. Since our starting
Hamiltonian is non Hermitian, the partners cannot be expected to
be inter-related through spersymmetry. On the contrary, it is
anticipated that they will be related by pseudo supersymmetry
\cite{mostafa-psusy}. It will also be shown that the pair of non
Hermitian Hamiltonians $H_{\pm}$ are related to a pair of
Hermitian ones $h_{\pm}$ through the same similarity
transformation $\rho$. Finally, we shall apply our formalism to
the entire class of shape-invariant potentials, where the
parameters of the partner potential are related to those of the
initial one through translation \cite{shape-invariant}. It is
worth mentioning here that we have been able to give a general
expression for finding the partner Hamiltonian $H_+$ in terms of
the parameters of $H_-$, for all the shape-invariant models
related through translation of parameters.

The plan of the work is as follows. In section 2, new non
Hermitian Hamiltonians $H_+$ are generated, which are isospectral
to the initial non Hermitian Hamiltonian $H_-$, except for the
ground state. It is observed further that both the initial non
Hermitian Hamiltonian $H_-$ and its partner $H_+$ so generated,
are pseudo Hermitian with respect to the same linear, invertible
operator $\eta$. The underlying symmetry between the partners
$H_{\pm}$ is studied in section 3. The formalism developed here is
actually applied to all the known classes of shape-invariant
models mentioned above, in section 4. Finally, Section 5 is kept
for conclusions and discussions.

\section{Theory}

For a better understanding of the topic and to make the paper
self-contained, we repeat certain equations from ref.
\cite{jpa-07} in the initial part of this section. To start with
we consider the generalized Swanson model
\begin{equation}
    H_- = {\cal{A}}^{\dagger} {\cal{A}} + \alpha {\cal{A}} ^2 +
    \beta {\cal{A}}^{\dagger \ 2}  \ ,  \qquad \alpha \neq \beta
\end{equation}
where $\alpha$ and $\beta$ are real, dimensionless constants, with
$\alpha \ \neq \ \beta $ for $H_-$ to be non Hermitian, and
${\cal{A}}^{\dagger} $ and ${\cal{A}}$ are generalized creation
and annihilation operators, given by
\begin{equation}\label{a}
    {\cal{A}} = \displaystyle \frac{d}{dx} + W(x) \qquad ,
    \qquad {\cal{A}} ^{\dagger} = \displaystyle - \frac{d}{dx} + W(x)
\end{equation}
Investigations in this field has revealed that for such non
Hermitian Hamiltonians to describe physical systems, they should
be necessarily $\eta$-pseudo Hermitian \cite{mostafa},
\begin{equation}\label{pseudoH}
    H_- ^{\dagger} \ = \ \eta H_- \eta ^{-1} \qquad , \qquad
    {\rm{i.e.}} \qquad
    H_-^{\dagger} \eta \ = \  \eta H_-
\end{equation}
where $\eta$ is a linear, invertible, Hermitian operator. This
requirement, along with the criterion for the wave functions to be
well behaved in the entire range, the parameters must obey certain
conditions \cite{jpa-07, swanson}, viz.,
\begin{equation}\label{parameters}
    \alpha ~+~ \beta ~ < ~ 1 \qquad , \qquad 4 \alpha \beta ~ < ~ 1
\end{equation}
With the explicit form of (\ref{a}) and some straightforward
algebra, the eigenvalue equation
\begin{equation}\label{evalue}
    H_- \psi ^{(-)} (x) = E \psi ^{(-)} (x)
\end{equation}
can be cast in the form \cite{jpa-07}
\begin{equation}\label{schro}
    \begin{array}{lcl}
    H_- \psi ^{(-)}
    &=& \displaystyle \left\{ - \left( 1 - \alpha - \beta
    \right) \left( \frac{d}{dx} - \frac{\alpha - \beta }{1
    - \alpha - \beta } \ W \right) ^2 +
    \frac{1 - 4 \alpha \beta }{\left( 1 - \alpha - \beta
    \right)} \ W^2
    -  \ W^{\prime} \right\} \psi ^{(-)} \\ \\
    &=& E \psi ^{(-)}
\end{array}
\end{equation}
To reduce eq. (\ref{schro}) to the well known Schr\"{o}dinger form
\begin{equation}\label{h-}
    h_- \ \phi ^{(-)}(x) ~=~ \displaystyle \left( - \frac{d^2}{dx ^2} \ + \ V_-(x)
    \right) \phi ^{(-)}(x) ~=~ \varepsilon \phi ^{(-)}(x)
\end{equation}
one has to apply a transformation of the form \cite{faria}
\begin{equation}\label{transform}
    \psi ^{(-)}(x) = \displaystyle \rho ^{-1}  \ \phi ^{(-)} (x)
\end{equation}
where
\begin{equation}\label{rho}
    \rho = \displaystyle e^{ - \mu \int W(x) ~ dx }
    \qquad , \qquad \rm{with} \ \ \ \mu = \displaystyle
    \frac{\alpha - \beta} {1 - \alpha - \beta}
    \ , \ \alpha ~+~ \beta ~ \neq ~ 1
\end{equation}
so that comparison between (\ref{schro}) and (\ref{h-}) gives
\begin{equation}\label{v-}
    \begin{array}{lcl}
        V_-(x) &=& \displaystyle \left( \frac{
        \sqrt{1 - 4 \alpha \beta} }{ 1 - \alpha - \beta } \ W(x)
    \right) ^2  ~-~ \frac{1}{\left( 1 - \alpha -
    \beta \right)} W^{\ \prime} (x) \\ \\
    \ \ \ \ \varepsilon  &=& \displaystyle
    \frac{E}{1 - \alpha - \beta}  \\
    \end{array}
\end{equation}
Thus, a quantum system described by a pseudo Hermitian Hamiltonian
$H_-$, is mapped to an equivalent system described by its
corresponding Hermitian counterpart $h_-$, with the help of a
similarity transformation $\rho$ \cite{jpa-07, mostafa2, jones},
\begin{equation}\label{similar}
    h_- = \rho H_- \rho ^{-1}
\end{equation}
We, now, take refuge in the formalism of supersymmetric quantum
mechanics (SUSYQM) \cite{susy}, or the equivalent intertwining
operator method \cite{intertwine}, to find an isospectral partner
of $h_-$. As is well known, $h_-$ can always be written in a
factorizable form as a product of a pair of linear differential
operators $ \tilde{A} \ , \tilde{A}^{\dagger} $, as
\begin{equation}\label{h-a}
    h _- = \displaystyle \tilde{A} ^{\dagger} \tilde{A}
    = \displaystyle -~ \frac{d^2}{dx^2} ~+~
    w^2 - w^{\prime}
\end{equation}
apart from some factorization energy $\epsilon$, where $ \tilde{A}
\ , \ \tilde{A} ^{\dagger} $ and $w(x)$ are given by
\begin{equation}\label{a-a}
    \tilde{A} ~=~ \displaystyle \frac{d}{dx} + w(x) \ , \qquad
    \tilde{A} ^{\dagger} ~=~ \displaystyle - \frac{d}{dx} + w(x)
    \ , \qquad w (x) = \displaystyle - \frac{d \ln \phi ^{-} _0 (x)}{dx}
\end{equation}
$\phi ^{-} _0$ being the ground state eigenfunction of $\tilde{A}
^{\dagger} \tilde{A}$ with energy $ \varepsilon_0 $. Thus $V_-
(x)$ in (\ref{v-}) can be identified with $ ( w^2 - w^{\prime} ) $
\begin{equation}\label{w-}
    V_- (x) = w^2 (x) - w^{\prime} (x)
\end{equation}
With the help of (\ref{v-}), the original eigenvalue equation
(\ref{schro}) may be written in a more compact form as
\begin{equation}\label{H1}
    \displaystyle H_- \psi ^{(-)} (x) =
    \left( 1 - \alpha - \beta \right)
    \left\{ - \left( \frac{d}{dx} - \frac{\alpha - \beta}{1 -
    \alpha - \beta} W(x) \right) ^2 + V_- (x) \right \} \psi
    ^{(-)} (x) = E \psi ^{(-)} (x)
\end{equation}
By the principles of SUSYQM, the hamiltonian $h_-$ is isospectral
to its partner Hamiltonian $h_+$ given by
\begin{equation}\label{h+a}
    h _+ = \displaystyle \tilde{A} \tilde{A} ^{\dagger}
    = \displaystyle -~ \frac{d^2}{dx^2} ~+~
    w^2 + w^{\prime}
\end{equation}
{\it i.e.,}
\begin{equation}\label{h+}
    h_+ \ \phi ^{(+)}(x) ~=~ \displaystyle \left( - \frac{d^2}{dx ^2} \ + \ V_+(x)
    \right) \phi ^{(+)}(x) ~=~ \varepsilon \phi ^{(+)}(x)
\end{equation}
where
\begin{equation}\label{v+}
    V_+ (x) = w^2 (x) + w^{\prime}(x)
\end{equation}
Let us now apply the inverse transformation of that given in
(\ref{transform}) to (\ref{h+}) above, i.e.,
\begin{equation}\label{inv-transform}
    \phi ^{(+)}(x) ~=~ \rho \  \psi ^{(+)} (x)
    ~=~ \displaystyle e^{- \mu \int W(x) dx } \psi ^{(+)} (x)
\end{equation}
After some straightforward algebra, equation (\ref{h+}) can be
written as
\begin{equation}\label{H2}
    \displaystyle H_+ \psi ^{(+)} =
    \displaystyle \left( 1 - \alpha - \beta \right)
    \left\{ - \left( \frac{d}{dx} - \frac{\alpha - \beta}{1 -
    \alpha - \beta} W(x) \right) ^2 + V_+ (x) \right \} \psi
    ^{(+)} = E \psi ^{(+)}
\end{equation}
Thus, $H_{\pm}$ are of the  same form, except for the explicit
form of $V_{\pm} (x)$. Evidently, both the initial Hamiltonian
$H_-$ as well as its partner $H_+$ are non Hermitian. Since
$h_{\pm}$ share identical energies, except for the ground state,
so should $H_{\pm}$, with the exception of the ground state. Thus,
applying the principles of SUSYQM, we obtain a non Hermitian
partner Hamiltonian $H_+$ of the initial one $H_-$, sharing
identical energies except for the ground state.

\subsection{Pseudo Hermiticity of $H_+$}

If one considers the inverse transformation (\ref{inv-transform}),
then it is easy to check that both the Hermitian Hamiltonian
$h_{\pm}$ and their non Hermitian counterparts $H_{\pm}$ are
related by the same similarity transformation as in
(\ref{similar}), i.e.,
\begin{equation}\label{similarity-2}
    H_{\pm} = \rho ^{-1} h_{\pm} \rho
\end{equation}
Additionally, simple algebra shows that both the non Hermitian
Hamiltonian $H_{\pm}$ are pseudo Hermitian with respect to the
same pseudo Hermiticity operator $\eta$
\begin{equation}\label{pseudo-2}
    H_{\pm} ^{\dagger} = \eta H_{\pm} \eta ^{-1} \qquad \qquad {\rm{i.e.}}
    \qquad \qquad H_{\pm} ^{\dagger} \eta = \eta H_{\pm}
\end{equation}
where $\rho$ and $\eta$ are inter-related through $ \rho =
\sqrt{\eta} \ $ \cite{jpa-07, jones}.

\vspace{.5cm}

\noindent It is interesting to study the behaviour of the wave
functions $\psi ^{(\pm)} (x)$. Since $H_{\pm}$ are $\eta$-pseudo
Hermitian, the wave functions should be normalized as $< \psi
^{(\pm)} \mid \eta \mid \psi ^{(\pm)} >$ \cite{mostafa}. With
$\eta = \rho ^2 $ and $ \psi ^{(\pm)} (x) = \rho ^{-1} \phi
^{(\pm)}(x) $, the above normalization condition reduces to the
conventional normalization of Hermitian quantum systems, viz., $ <
\phi ^{(\pm)} \mid \phi ^{(\pm)} > $, easily available in standard
text books of quantum mechanics for the shape-invariant potentials
considered here \cite{susy}.

\vspace{1cm}

\section{Underlying symmetry between the partners
$H_{\pm}$}

To explore the underlying symmetry between the isospectral
partners $H_{\pm}$, we start with their Hermitian counterparts
$h_{\pm}$. Now, $h_{\pm}$ form a pair of supersymmetric partners,
with super Hamiltonian
\begin{equation}\label{super-h}
    h = \left(%
    \begin{array}{cc}
        h_- & 0 \\
        0 & h_+  \\
    \end{array}
    \right)
\end{equation}
and generated by supercharges
\begin{equation}\label{charge}
    q = \left(%
    \begin{array}{cc}
    0 & \tilde{A} ^{\dagger} \\
    0 & 0 \\
    \end{array}
    \right), \ \ \ \ \ \ \ \ \ \ \ \ \ \ q^{\dagger} =
    \left(%
    \begin{array}{cc}
    0 & 0 \\
    \tilde{A}  & 0 \\
    \end{array}
    \right)
\end{equation}
so that
\begin{equation}\label{h-charges}
    h ~=~ \displaystyle \left \{ q^{\dagger} , q \right \}
\end{equation}
To establish the symmetry relation between $H_{\pm}$, we return to
the similarity transformation between the original non Hermitian
Hamiltonian $H_-$ and its Hermitian mapping $h_-$, viz.,
$$ H_- = \rho ^{-1} h_- \rho  $$
If one defines two operators $D_{\pm}$ {\footnote{In case the
operators $D_{\pm}$ are defined by taking the negative square root
of $(1 - \alpha - \beta)$, then the new operators $D_{\pm} ^{new}
$ so formed are related to $D_{\pm}$ through a constant phase,
viz., $D_{\pm} ^{new} = e^{\pm i \pi } D_{\pm} $. This introduces
no change in the expression for $H_{\pm}$.}} as
\begin{equation}\label{dd}
    D_+ = \displaystyle \left
    ( \sqrt{ 1 - \alpha - \beta } \right) \
    \rho ^{-1} \tilde{A} ^{\dagger} \rho  \qquad \qquad
    D_- = \displaystyle \left( \sqrt{ 1 - \alpha - \beta }
    \right) \ \rho ^{-1} \tilde{A} \ \rho
\end{equation}
then the isospectral Hamiltonians, $H_{\pm}$ , can be written in
terms of these operators as
\begin{equation}\label{h12}
    H_- = D_+ D_- \qquad, \qquad \qquad \qquad H_+ = D_- D_+
\end{equation}
so that $D_{\pm}$ play the role of intertwining operators for
$H_{\pm}$
\begin{equation}\label{intertwine}
    D_- H_- = H_+ D_- \qquad \qquad , \qquad \qquad
    H_- D_+ = D_+ H_+
\end{equation}
With the help of (\ref{a}) and (\ref{dd}), $D_{\pm}$ can be
written in the explicit form
\begin{equation}\label{d-w}
    \begin{array}{lcl}
    D_+ = \displaystyle \left( \sqrt{ 1 - \alpha - \beta }
    \right) \ \left\{- \frac{d}{dx} + \mu W (x) + w(x)
    \right\} \\ \\
    D_- = \displaystyle \left( \sqrt{ 1 - \alpha - \beta }
    \right) \ \left\{ \frac{d}{dx} - \mu W (x) + w(x) \right\}
    \end{array}
\end{equation}
It is worth noting here that the functions $W(x)$ and $w(x)$
appearing in the explicit form of $D_{\pm}$ are not independent.
Instead, they are related to each other by equations (\ref{v-})
and (\ref{w-}), i.e.
\begin{equation}\label{v-w}
    w^2 (x) - w^{\prime} (x) ~=~ \displaystyle \left( \frac{
        \sqrt{1 - 4 \alpha \beta} }{ 1 - \alpha - \beta } \ W(x)
    \right) ^2  ~-~ \frac{1}{\left( 1 - \alpha -
    \beta \right)} W^{\ \prime} (x)
\end{equation}
Since the isospectral partner Hamiltonians $H_{\pm}$ are pseudo
Hermitian, we expect them to be embedded in the framework of
pseudo supersymmetry \cite{mostafa-psusy}. Straightforward algebra
shows that the operators $D_{\pm}$ are pseudo-adjoint of one
another
\begin{equation}\label{pseudo-adjoint}
    \displaystyle \left( D_+ \right) ^{\sharp} ~=~ \eta ^{-1} \left( D_+
    \right) ^{\dagger} \eta ~=~ \eta ^{-1} \left( \rho \tilde{A} \ \rho ^{-1}
    \right) \eta ~=~ \rho ^{-1} \tilde{A} \rho ~=~ D_-
\end{equation}
If we define two operators $Q$ and $Q ^\sharp$ as
\begin{equation}\label{pseudo-charge}
    Q = \left(%
    \begin{array}{cc}
    0 & D_+ \\
    0 & 0 \\
    \end{array}
    \right), \ \ \ \ \ \ \ \ \ \ \ \ \ \ Q ^{\sharp} = \eta ^{-1}
    Q ^{\dagger} \eta = \left(%
    \begin{array}{cc}
    0 & 0 \\
    D_-  & 0 \\
    \end{array}
    \right) =
    \left(%
    \begin{array}{cc}
    0 & 0 \\
    \left( D_+ \right) ^ {\sharp}  & 0 \\
    \end{array}
    \right)
\end{equation}
and construct a new Hamiltonian ${\cal{H}}$ from the partners
$H_{\pm}$ as
\begin{equation}\label{pseudosuper-H}
    {\cal{H}} = \left(%
    \begin{array}{cc}
        H_- & 0 \\
        0 & H_+  \\
    \end{array}
    \right)
\end{equation}
then it is easy to observe that
\begin{equation}\label{H-qq}
    {\cal{H}} ~=~ \displaystyle \left \{ Q ^{\sharp} , Q \right \}
\end{equation}
Additionally,
\begin{equation}\label{nilpotent}
    \displaystyle \left \{ Q, Q \right \} ~=~
    \left \{ Q^{\sharp}, Q^{\sharp} \right \} ~=~ 0
\end{equation}
Thus we obtain the standard pseudo super algebra of non Hermitian
supersymmetry \cite{mostafa-psusy}, with the operators $Q$ and $Q
^\sharp $ playing the role of pseudo super charges, the
anticommutator of which gives the pseudo super Hamiltonian
${\cal{H}}$. Interestingly, though it may not be possible (in
general) to express the new Hamiltonian $H_+$ in terms of the
generalized annihilation and creation operators ${\cal{A}}$ and
${\cal{A}}^{\dagger}$, nevertheless, the isospectral partners
$H_{\pm}$ can be shown to be related by pseudo supersymmetry.
Furthermore, it is also observed that the super charges $q \ , \
q^{\dagger} $ of conventional supersymmetry are related to the
pseudo supercharges $Q \ , \ Q^{\sharp}$ of pseudo supersymmetry
through
\begin{equation}\label{qq}
    Q = \rho ^{-1} \ q \ \rho
\end{equation}
This follows from the similarity mapping between the non Hermitian
Hamiltonians $H_{\pm}$ and their respective Hermitian counterparts
$h_{\pm}$. We shall devote the next section to construct some non
Hermitian Hamiltonians as isospectral partners of the generalized
Swanson models based on those shape-invariant potentials where the
parameters are related to each other by translation ($ a_2 = a_1 +
\lambda $) \cite{susy, shape-invariant}.

\section{Models based on shape-invariant potentials}

For our formalism to be applicable to specific models, one needs
to solve the highly non-trivial Ricatti equation (\ref{v-w}). This
demands certain restrictions on the forms of $W(x)$ and $w(x)$. If
one wants to map a certain type of potential (say Harmonic
oscillator) to a different type (say e.g., P\"{o}schl-Teller or
Rosen-Morse), while going from the non Hermitian to the Hermitian
picture, the corresponding Ricatti equation cannot be solved
analytically (or, at least in an obvious way). For the shape
invariant class, the function $w(x)$ consists of two parts,
denoted by $f(x)$ and $g(x)$, i.e.
\begin{equation}\label{w1-fg}
    w(x) = \lambda _1 f(x) ~+~ \delta _1 g(x) \qquad ,
    \qquad {\rm{with}} \ \ \lambda _1 \ ,
    \ \delta _1 \ \ {\rm{constants}}
\end{equation}
For reasons given in the beginning of this section, the function
$W(x)$ used in the construction of generalized annihilation and
creation operators ${\cal{A}}$ and ${\cal{A}}^{\dagger}$ in
(\ref{a}) is assumed to be of the same form as $w(x)$ :
\begin{equation}\label{w2-fg}
    W(x) = \lambda _2 f(x) ~+~ \delta _2 g(x) \qquad ,
    \qquad {\rm{with}} \ \ \lambda _2 \ ,
    \ \delta _2 \ \ {\rm{constants}}
\end{equation}
Our aim is to write $V_-(x)$ in terms of $w^2(x) - w^{\prime}(x)$.
It is already shown that $W(x)$ and $w(x)$ are inter-related
through (\ref{v-w}). Substituting (\ref{w1-fg}) and (\ref{w2-fg})
in (\ref{v-w}), the expression takes the explicit form
\begin{equation}\label{ww}
    \begin{array}{lll}
        & & \displaystyle \frac{1-4 \alpha \beta}{ \left( 1 - \alpha -
        \beta \right) ^2 } \left\{ \lambda _2 ^2 f^2  + \delta _2
        ^2 g^2 + 2 \lambda _2 \delta _2 fg \right\}
        - \frac{1}{1 - \alpha - \beta } \left( \lambda _2 f ^{\ \prime}
        + \delta _2 g^{\ \prime} \right) \\ \\
        & & \displaystyle = \lambda _1 ^2 f^2  + \delta _1
        ^2 g^2 + 2 \lambda _1 \delta _1 fg -
        \lambda _1 f ^{\ \prime} - \delta _1 g^{\ \prime}
    \end{array}
\end{equation}
This general expression relates the unknown parameters $ \lambda
_1, \delta _1 $ in terms of the known ones $ \lambda _2, \delta _2
$, for all shape invariant potentials where the parameters of the
original potential and its partner are related to each other by
translation. This enables one to write the partner potential
$V_+(x)$, and hence the partner Hamiltonian $H_+$, in terms of the
parameters of the starting Hamiltonian $H_-$.  Now these
shape-invariant models can be further classified under different
categories, depending on the particular forms of $f(x)$ and
$g(x)$. We shall explore these in further detail in the next few
subsections.

\vspace{.5cm}

\subsection{Case 1 : $g(x) = {\rm{constant}} \ , \
\ f^2 (x) = c_1 f^{\prime} (x) + c_2 \ , $  \\
with $c_1, \ c_2 $ constants}

\vspace{.3cm}

\noindent In this subsection, we shall study the models based on
the following potentials : \\ \\
{\bf 1. Rosen Morse I (trigonometric) potential}
\begin{equation}\label{rm1}
    V(x) = \displaystyle a(a -1)\csc ^2 x + 2b \cot x
    - a ^2 + \frac{b ^2}{a ^2} \ \ \ \ ,  \qquad 0
    \leq x \leq \pi
\end{equation}
\begin{equation}\label{w-rm1}
    {\rm{with}} \ \ \ W(x) = \displaystyle - a_2 \cot  x - \frac{b_2}{a_2}
    \qquad , \qquad  a_2 > 0 \ , \ b_2 > 0
\end{equation}
{\bf 2. Rosen Morse II (hyperbolic) potential}
\begin{equation}\label{rm2}
    V(x) = \displaystyle - a(a +1) \ {\rm{sech}} ^2 x + 2b \tanh x
    + a ^2 + \frac{b ^2}{a ^2} \ , \qquad
    b < a ^2 \ \ \ \ ,  \qquad - \infty
    \leq x \leq \infty
\end{equation}
\begin{equation}\label{w-rm2}
    {\rm{with}} \ \ \ W(x) = \displaystyle  a_2 \tanh  x + \frac{b_2}{a_2}
    \qquad , \qquad  a_2 > 0 \ , \ b_2 > 0
\end{equation}
{\bf 3. Eckart potential}
\begin{equation}\label{eckart}
    V(x) = \displaystyle  a(a -1) \ {\rm{cosech }} ^2 x
    - 2b \coth x
    + a ^2 + \frac{b ^2}{a ^2} \ , \qquad
    b > a ^2 \ \ \ \ ,  \qquad 0
    \leq x \leq \infty
\end{equation}
\begin{equation}\label{w-eck}
    {\rm{with}} \ \ \ W(x) = \displaystyle - a_2 \coth  x + \frac{b_2}{a_2}
    \qquad , \qquad  a_2 > 0 \ , \ b_2 > 0
\end{equation}

\vspace{.2cm}

\noindent For the sake of convenience, we put $g(x) = 1$. This
simplifies (\ref{ww}) to
\begin{equation}\label{ww-1}
    \begin{array}{lll}
        & & \displaystyle \left( \lambda _1 ^2 c_1 - \lambda _1 \right)
        f^{\prime} (x) + 2 \lambda _1 \delta _1 f(x) + \lambda _1 ^2
        c_2 + \delta _1 ^2 \\ \\
        & & \displaystyle = \left( \frac{(1 - 4 \alpha \beta)
        \lambda_2 ^2 c_1}{( 1 - \alpha - \beta )^2} - \frac{\lambda _2}{1
        - \alpha - \beta)} \right) f^{\prime} (x) +
        \frac{2 \lambda _2 \delta _2
        (1 - 4 \alpha \beta)}{(1 - \alpha - \beta )^2} f(x) \\ \\
        & & \displaystyle \ \ \ \ +
        \frac{1 - 4 \alpha \beta}{(1 - \alpha - \beta )^2} \left(
        \lambda _2 ^2 c_2 + \delta _2 ^2 \right)
    \end{array}
\end{equation}
Equating like terms on both sides, the unknown parameters $
\lambda _1,\delta _1 $ are expressed in terms of the known ones $
\lambda _2,\delta _2 $ through the following :
\begin{equation}\label{ww1-lambda}
    \displaystyle \lambda _1 ^2 c_1 - \lambda _1 =
    \frac{\lambda _2 ^2 c_1 (1 - 4 \alpha \beta)}{
    (1 - \alpha - \beta )^2} - \frac{\lambda _2}{1 - \alpha - \beta }
\end{equation}
or, more explicitly,
\begin{equation}\label{lambda-1}
    \lambda _1 = \displaystyle \frac{1 \pm \sqrt{1
    + 4 \sigma _-}}{2 c_1}
\end{equation}
where
\begin{equation}\label{sigma-1}
    \sigma _- = \displaystyle \frac{\lambda _2 ^2 c_1 (1 - 4 \alpha \beta )}{
    (1 - \alpha - \beta )^2} - \frac{\lambda _2}{1 - \alpha - \beta }
\end{equation}
and
\begin{equation}\label{ww1-delta}
    \delta _1 = \frac{\lambda _2 \delta _2 }{\lambda _1}
    \frac{1 - 4 \alpha \beta}{1 - \alpha - \beta}
\end{equation}
Since $\lambda _1 $ and $ \lambda _2$ should be of the same sign,
only the positive sign is allowed in the expression for $ \lambda
_1 $ in (\ref{lambda-1}). The pseudo supersymmetric partners
$H_{\pm}$, expressed as,
\begin{equation}\label{partner}
    H_{\pm} \psi ^{(\pm)} (x) = E ^{(\pm)} \psi ^{(\pm)}
    (x)
\end{equation}
or, more explicitly,
\begin{equation}\label{H+-}
    H_{\pm}(x) = \displaystyle (1 - \alpha - \beta)
    \left\{ - \left( \frac{d}{dx} - \frac{\alpha - \beta}{1 -
    \alpha - \beta} W(x) \right) ^2 + V_{\pm} (x) \right\}
\end{equation}
have identical energies except for the ground state, with $V_{\pm}
(x)$ for this class of potentials reducing to
\begin{equation}\label{v-pm}
    V_{\pm}(x) = \displaystyle
    \left( \lambda _1 ^2 c_1 \pm \lambda _1 \right) f^{\prime} (x)
    + 2 \lambda_1 \delta _1 f(x) + \delta _1 ^2 + c_2 \lambda _1
    ^2
\end{equation}
The potentials falling in this category are listed below. In each
case the form of $w(x)$ is similar to that of $W(x)$, with $a_2 $
and $b_2$ being replaced by $a_1$ and $b_1$. The unknown
parameters $a_1$ and $b_1$ are obtained in terms of the known ones
$a_2$ and $b_2$ from expressions (\ref{lambda-1}), (\ref{sigma-1})
and (\ref{ww1-delta}). It can be shown that the eigen energies of
the positive and the negative sector are related through
\begin{equation}\label{e+}
    E_n ^{(+)} = E_{n+1} ^{(-)} \qquad ,
    \qquad {\rm{with}} \qquad
    E ^{(\pm)} = (1 - \alpha - \beta) \varepsilon ^{(\pm)},
    \qquad n = 0, 1, 2, \cdots
\end{equation}
The partner potentials $V_{\pm} (x)$ are given in Table 1 while
the (unnormalized) solutions of the original Hamiltonian $H_-$ are
given in Table 2. The solutions of its partner $H_+$ can be
obtained by applying the transformation
\begin{equation}\label{psi+}
    \psi _n ^{(+)}(x) ~=~ \rho ^{-1} \  \phi _n ^{(+)} (x)
\end{equation}
where $ \phi _n ^{(+)} $ are the solutions of the supersymmetric
partner Hamiltonian $h_+$. In the expression for $ \psi _n ^{(-)}
$,  the different parameters stand for
\begin{equation}
    \displaystyle
    \mu _1 = a_1 \mu = a_1 \frac{\alpha - \beta}{ 1 - \alpha - \beta }
    \qquad , \qquad \mu _2 = \frac{b_1}{a_1} \mu
    = \frac{b_1 (\alpha - \beta)}{a_1 ( 1 - \alpha - \beta)}
\end{equation}
It is evident from the explicit expressions for $\psi _n ^{(-)}
(x) $ that for its well-defined behaviour, $ \alpha $ and $ \beta
$ must obey additional constraints; e.g., for the Rosen Morse II
and Eckart models,
\begin{equation}
    \alpha < \beta
\end{equation}
while the Rosen Morse I model requires
\begin{equation}\
    a_1 + n + \mu _2 > 0 \qquad , \qquad \displaystyle
    \frac{b_1}{a_1 + n} < \mu _1
\end{equation}
which , in turn, implies
\begin{equation}
    \alpha > \beta
\end{equation}

\pagebreak


\noindent  {\bf Table 1 :}

\vspace{.2cm}

\noindent
\begin{tabular}{|c|c|c|c|c|}
  \hline
  Model &  $f(x)$ & $W(x)$ & $V_{\pm} (x) $ & $\varepsilon^{(-)} _n $ \\
  \hline
    & & & & \\
    Rosen Morse I & & &
    $ \displaystyle - \left( a_2 ^2  - \frac{b_2 ^2}{a_2 ^2}
    \right) \frac{1 - 4 \alpha \beta}{( 1 - \alpha - \beta )^2} $  &
    $ \displaystyle - \left( a_2 ^2  - \frac{b_2 ^2}{a_2 ^2}
    \right) \frac{1 - 4 \alpha \beta}{( 1 - \alpha - \beta )^2} $ \\
    & & & & \\
    $\lambda_2 = - a_2 $ & $ \cot x$ &
    $ - a_2 \cot x $ & $ + a_1(a_1 \pm 1) \csc ^2 x $ &
    $ +(a_1 + n)^2  $ \\
    & & & & \\
    $\displaystyle \delta _2 = - \frac{b_2}{a_2} $
    & & $ \displaystyle - \frac{b_2}{a_2} $ &
    $ \displaystyle + 2 b_2 \frac{1 - 4 \alpha \beta}{ (1 - \alpha - \beta )^2}
    \cot x  $ &
    $ \displaystyle - \frac{b_1 ^2}{( a_1 + n )^2} $ \\
    & & & & \\
    $ c_1 = c_2 = -1 $  & & & & $ n = 0, 1, 2, \cdots $ \\
    \hline
    & & & &  \\
    Rosen Morse II & & & $ \displaystyle \left( a_2 ^2  + \frac{b_2 ^2}{a_2 ^2}
    \right) \frac{1 - 4 \alpha \beta}{( 1 - \alpha - \beta )^2} $  &
    $ \displaystyle \left( a_2 ^2  + \frac{b_2 ^2}{a_2 ^2}
    \right) \frac{1 - 4 \alpha \beta}{( 1 - \alpha - \beta )^2} $ \\
    & & & & \\
    $ \lambda _2 = a_2 $ & $\tanh x$ &
    $ a_2 \tanh x $ & $ - a_1 (a_1 \pm 1) \ {\rm{sech}} ^2 x $ &
    $ - (a_1  - n )^2  $ \\
    & & & & \\
    $\displaystyle \delta _2 = \frac{b_2}{a_2}$
    & & $ + \displaystyle \frac{b_2}{a_2} $ &
    $ \displaystyle + 2 b_2
    \frac{1 - 4 \alpha \beta}{(1 - \alpha - \beta)^2}
    \tanh x $ &
    $ \displaystyle - \frac{b_1 ^2}{(a_1 - n ) ^2} $ \\
    & & & & \\
    $c_1 = - 1, c_2 = 1$ & & & & $ n = 0, 1, 2, \cdots  < a_1$ \\
    \hline
    & & & & \\
    Eckart & & & $ \displaystyle \left( a_2 ^2  + \frac{b_2 ^2}{a_2 ^2}
    \right) \frac{1 - 4 \alpha \beta}{( 1 - \alpha - \beta )^2} $
    & $ \displaystyle \left( a_2 ^2  + \frac{b_2 ^2}{a_2 ^2}
    \right) \frac{1 - 4 \alpha \beta}{( 1 - \alpha - \beta )^2} $ \\
    & & & & \\
    $ \lambda _2 = - a_2 $ & $\coth x$ &
    $ -a_2 \coth x $ & $ + a_1(a_1 \pm 1) \ {\rm{csch}} ^2 x $ &
    $ -(a_1 + n)^2  $ \\
    & & & & \\
    $\displaystyle \delta _2 = \frac{b_2}{a_2} $ & &
    $ + \displaystyle \frac{b_2}{a_2} $ &
    $ \displaystyle - 2 b_2
    \frac{1 - 4 \alpha \beta}{(1 - \alpha - \beta)^2} \coth x  $ &
    $ \displaystyle - \frac{b_1 ^2}{(a_1 + n )^2} $ \\
    & & & & \\ $c_1 = -1, c_2 = 1 $ & &
    & & $ n = 0, 1, 2, \cdots $ \\
    \hline
\end{tabular}

\vspace{.5cm}

\pagebreak

\noindent  {\bf Table 2 :}

\vspace{.2cm}

\noindent
\begin{tabular}{|c|c|c|c|}
  \hline
  Model & $y$ &  $s_{\pm} $ & $\psi _n ^{(-)}$ \\
  \hline
  & & & \\
  Rosen Morse I & $ i \cot x $ & $ \displaystyle
  - a_1 - n \pm i \frac{b_1}{a_1 + n} $ & $ \displaystyle
  e^{ \left( \frac{b_1}{a_1 +n} - \mu _1\right) x}
  \sin ^{a_1 + n + \mu_2} x \
  P_n ^{(s_+,s_-)} (y) $ \\ \hline
  & & & \\
  Rosen Morse II & $\tanh x $ & $ \displaystyle
  a_1 - n \pm \frac{b_1}{a_1 - n} $ & $ \displaystyle
  (1 - y)^{\frac{s_+ - \mu _1 }{2}}
  (1 + y)^{\frac{s_- - \mu _1 }{2}} e^{\mu _2 x}
  P_n ^{(s_+,s_-)} (y) $  \\ \hline
  & & & \\
  Eckart & $ \coth x $ & $ \displaystyle \pm
  \frac{b_1}{a_1 + n } - n -  a_1 $ &  $ \displaystyle
  (y - 1)^{\frac{s_+ + \mu _1 }{2}}
  (y + 1)^{\frac{s_- + \mu _1 }{2}} e^{\mu _2 x}
  P_n ^{(s_+,s_-)} (y) $  \\
  \hline
\end{tabular}

\vspace{.2cm}

\noindent

\vspace{1cm}

\subsection{Case 2 : $f^2(x) = c_1 + c_2 g^2 (x)  \ , \
f^{\prime} (x) = c_3 g^2(x)  \  , \ g^{\prime} (x) = c_4 f(x) g(x)
$
\\ with $ \ c_1 , \ c_2 , \ c_3  , \ c_4 $ constants}

\vspace{.3cm}

\noindent The models falling in this category are based on the : \\ \\
{\bf 1. Scarf I (trigonometric) potential}
\begin{equation}\label{scarf1}
    V(x)= k_1 \tan ^2 x - k_2 \sec x \ \tan x \qquad , \qquad
    \displaystyle - \frac{\pi}{2} \leq x \leq \frac{\pi}{2}
\end{equation}
\begin{equation}\label{w-scarf1}
    {\rm{with}} \ \ \
    W(x) = \displaystyle \lambda_2 \tan x  - \delta _2 \sec x
\end{equation}
{\bf 2. Scarf II (hyperbolic) potential}
\begin{equation}\label{scarf2}
    V(x)= k_1 \ {\rm{sech}}^2 x + k_2 \ {\rm{sech}} \ x \ \tanh x
    \qquad , \qquad - \infty \leq x \leq \infty
\end{equation}
\begin{equation}\label{w-scarf2}
    {\rm{with}} \ \ \
    W(x) = \displaystyle \lambda_2 \tanh x
    + \delta _2 {\rm{sech}} \ x
\end{equation}
{\bf 3. P\"{o}schl Teller potential}
\begin{equation}\label{poschl}
    V(x)= k_1 \ {\rm{cosech}}^2 x - k_2 \ \coth x \ {\rm{cosech}} \ x
    \qquad , \qquad 0 \leq x \leq \infty
\end{equation}
\begin{equation}\label{w-poschl}
    {\rm{with}} \ \ \
    W(x) = \displaystyle \lambda_2 \tanh x  - \delta _2
    {\rm{cosech}} \ x \qquad , \qquad \lambda_2 < \delta _2
\end{equation}

\vspace{.2cm}

\noindent Thus (\ref{ww}) gets simplified to
\begin{equation}\label{ww-2}
    \begin{array}{lll}
        & & \displaystyle \lambda _1 ^2 c_1 + \left( \lambda _1 ^2 c_2
        + \delta _1 ^2 - \lambda _1 c_3 \right) g^2 (x)
        + \left( 2 \lambda _1 \delta _1 - \delta _1 c_4 \right) f(x) g(x)
        + \lambda _1 ^2 c_1  \\ \\
        & & \displaystyle = \left[ \frac{(1 - 4 \alpha \beta)
        }{( 1 - \alpha - \beta )^2} \left( \lambda _2 ^2 c_2 +
        \delta _2 ^2 \right)- \frac{\lambda _2 c_3 }{1
        - \alpha - \beta)} \right] g^2 (x) +
        \frac{\lambda _2 ^2 c_1 (1 - 4 \alpha \beta)}{(1 - \alpha - \beta )^2} \\ \\
        & & \displaystyle \ \ \ \ +
        \left[ 2 \lambda _2 \delta _2
        \frac{1 - 4 \alpha \beta}{(1 - \alpha - \beta )^2}
        - \frac{\delta _2 c_4}{ 1 - \alpha - \beta} \right]
        f(x)g(x)
    \end{array}
\end{equation}
Equating like terms on both sides, the unknown parameters $
\lambda _1,\delta _1 $ are obtained by solving the following two
coupled equations simultaneously :
\begin{equation}\label{couple-1}
    \displaystyle \delta _1 ^2 + \lambda _1 ^2 c_2 - \lambda _1 c_3 =
    \left( \lambda _2 ^2 c_2 + \delta _2 ^2 \right)
    \frac{ (1 - 4 \alpha \beta)}{(1 - \alpha - \beta )^2} -
    \frac{\lambda _2 c_3 }{1 - \alpha - \beta }
\end{equation}
\begin{equation}\label{couple-2}
    2 \lambda _1 \delta _1 - \delta _1 c_4 = 2 \lambda _2 \delta _2
        \frac{1 - 4 \alpha \beta}{(1 - \alpha - \beta )^2}
        - \frac{\delta _2 c_4}{ 1 - \alpha - \beta}
\end{equation}
Once again, the pseudo supersymmetric partner Hamiltonians
$H_{\pm}$, given by (\ref{H+-}), have identical energies except
for the ground state, with $V_{\pm} (x)$ for this class of
potentials assuming the form
\begin{equation}\label{v-pm2}
    V_{\pm}(x) = \displaystyle
    \lambda _1 ^2 c_1 + \left( \lambda _1 ^2 c_2
        + \delta _1 ^2 \pm \lambda _1 c_3 \right) g^2 (x)
        + \left( 2 \lambda _1 \delta _1 \pm \delta _1 c_4 \right) f(x) g(x)
        + \lambda _2 ^2 c_1 \frac{1 - 4 \alpha \beta}{( 1 - \alpha
        - \beta )^2}
\end{equation}

\vspace{1cm}

\noindent In each of the cases of the three different models in
this category, the form of $w(x)$ is similar to that of $W(x)$,
with $\lambda_2 $ and $\delta_2$ being replaced by $\lambda_1$ and
$\delta_1$. The unknown parameters $\lambda_1$ and $\delta_1$ are
obtained in terms of the known ones $\lambda_2$ and $\delta_2$
from expressions (\ref{lambda-1}), (\ref{sigma-1}) and
(\ref{ww1-delta}). The pseudo supersymmetric partner Hamiltonians
of the form given in (\ref{H+-}), have energies $E^{(\pm)}$,
related to $\varepsilon ^{(\pm)}$ through (\ref{e+}). The partner
potentials $V_{\pm} (x)$ are given in Table 3 while the solutions
are given in Table 4, with
\begin{equation}\label{mu-12}
    \mu _1 = \lambda_2 \mu \qquad , \qquad
    \mu _2 = \delta _2 \mu
\end{equation}
From the explicit expressions for the solutions, it is evident
that well defined behaviour is assured only when the parameters
satisfy additional constraints. For example, for the P\"{o}schl
Teller model, this condition reduces to
$$ \mu _2 < 0 \qquad {\rm{i.e.}} \qquad
\alpha < \beta $$


\pagebreak

\noindent  {\bf Table 3 :}

\vspace{.2cm}

\noindent
\begin{tabular}{|c|c|c|c|c|}
  \hline
  Model &  $f(x)$ & $g(x) $ & $V_{\pm} (x) $ & $\varepsilon^{(-)} _n $ \\
  \hline
    & & & &  \\
    Scarf I & & &
    $ \displaystyle  - \lambda_2 ^2
    \frac{1 - 4 \alpha \beta}{( 1 - \alpha - \beta )^2}
    - \lambda _1 ^2 $  &
    $ \displaystyle  - \lambda_2 ^2
    \frac{1 - 4 \alpha \beta}{( 1 - \alpha - \beta )^2} $ \\
    & & & &  \\
    $c_1 = -1  , \ c_2 = 1$ & $  \tan x$ & $ - \sec x $
    & $ + \left( \delta_1 ^2 +
    \lambda_1 ^2 \pm \lambda_1 \right) \sec ^2 x $ &
    $ - \lambda_1 ^2 + ( \lambda_1 + n)^2  $ \\
    & & & & \\
    $c_3 = 1 , \ c_4 = 1  $
    & & &
    $ \displaystyle - \delta _1 \left( 2 \lambda _1 \pm 1
    \right) \sec x \ \tan x $ &
    $ n = 0, 1, 2, \cdots $ \\
    & & & &  \\  \hline
    & & & &  \\
    Scarf II  & & & $ \displaystyle  \lambda_2 ^2
    \frac{1 - 4 \alpha \beta}{( 1 - \alpha - \beta )^2}
    + \lambda _1 ^2  $  &
    $ \displaystyle \lambda_2 ^2
    \frac{1 - 4 \alpha \beta}{( 1 - \alpha - \beta )^2} $ \\
    & & & &  \\
    $ c_1 = 1 \ , \ c_2 = -1 $ & $\tanh x$ & ${\rm{sech}} \ x$ &
    $ \displaystyle + \left( \delta _1 ^2 - \lambda _1 ^2 \pm
    \lambda _1 \right) \ {\rm{sech}} ^2 x $ &
    $ + \lambda _1 ^2 - (\lambda_1  - n )^2  $ \\
    & & & &  \\
    $c_3 = 1 \ , \ c_4 = -1$
    & & &
    $ + \delta _1 ( 2 \lambda _1 \mp 1 ) \ {\rm{sech}} \ x \
    \tanh x $ &
    $ n = 0, 1, 2, \cdots < \lambda _1 $ \\
    & & & &  \\  \hline
    & & & & \\
    P\"{o}schl-Teller & & & $ \displaystyle \lambda_2 ^2
    \frac{1 - 4 \alpha \beta}{( 1 - \alpha - \beta )^2}
    + \lambda _1 ^2 $
    & $ \displaystyle \lambda_2 ^2
    \frac{1 - 4 \alpha \beta}{( 1 - \alpha - \beta )^2} $ \\
    & & & & \\
    $ c_1 = 1 \ , \ c_2 = 1 $ & $\coth x$ & $ - {\rm{csch}} \ x $
    & $ + ( \delta_1 ^2 + \lambda_1 ^2  \pm \lambda_1) \ {\rm{csch}} ^2 x $ &
    $ + \lambda_1 ^2 -(\lambda_1 - n)^2  $ \\
    & & & & \\
    $c_3 = -1, c_4 = -1 $ & & &
    $ \displaystyle - \delta_1 ( 2 \lambda_1 \mp 1)
    \ {\rm{csch}} \ x \ \coth x  $ &
    $ n = 0, 1, 2, \cdots < \lambda _1 $ \\
    & & & &  \\    \hline
\end{tabular}

\vspace{1cm}


\noindent  {\bf Table 4 :}

\vspace{.2cm}

\noindent
\begin{tabular}{|c|c|c|c|}
  \hline
  Model & $y$ &  $s_{\pm} $ & $\psi _n ^{(-)}$ \\
  \hline
  & & & \\
  Scarf I & $ \sin x $ & $ \displaystyle
  \lambda_1  \pm \delta_1 - \frac{1}{2} $ & $ \displaystyle
  (\sec x + \tan x)^{- \mu _2}
  (1-y)^{\frac{\lambda_1 - \delta_1 - \mu _1}{2}}
  (1+y)^{\frac{\lambda_1 + \delta_1 - \mu _1}{2}}
  P_n ^{(s_-,s_+)} (y) $ \\ \hline
  & & & \\
  Scarf II & $ \sinh x $ & $ \displaystyle
  \pm i \delta_1 - \lambda_1  - \frac{1}{2} $ & $ \displaystyle
  (1 + y^2)^{\frac{\mu _1 - \lambda_1 }{2}} e^{(\mu _2 - \delta_1 )
  {\rm{tan}} ^{-1} y }
  P_n ^{(s_+ ,s_- )} (y) $  \\ \hline
  & & & \\
  P\"{o}schl-Teller & $ \cosh x $ & $ \displaystyle
  \pm \delta_1 - \lambda_1  - \frac{1}{2} $ &  $ \displaystyle
  (y - 1)^{\frac{\delta_1 - \lambda_1  + \mu _1 }{2}}
  (y + 1)^{\frac{- \delta_1 - \lambda_1 + \mu _1 }{2}} e^{\mu _2 x}
  P_n ^{(s_+ , s_- )} (y) $  \\
  \hline
\end{tabular}

\vspace{1cm}

\subsection{Case 3 : $g(x) = 1 \ , \ {\rm{and}}
\ f^{\prime} (x) = k f (x) \ , \ {\rm{with}} \ k = - 1 $ }

\vspace{.3cm}

\noindent The {\bf Morse potential}, given by
\begin{equation}\label{morse}
    V(x) = \displaystyle a_1 ^2 + b_1 ^2 \exp (-2x)
    - b_1 (2 a_1 + 1) \exp (-x) \qquad , \qquad
    - \infty \leq x \leq \infty
\end{equation}
belongs to this class of potentials, with
\begin{equation}\label{w-morse}
    W(x) = a_2 - b_2 \exp (-x)
\end{equation}
Thus, for this particular model, $ \lambda _2 = -b_2 \ , \ \delta
_2 = a_2 $, with  $f(x) = \exp (-x) $, so that equation (\ref{ww})
reduces to
\begin{equation}\label{lambda-morse}
    \displaystyle b_1 = b_2 \frac{\sqrt{1 - 4 \alpha
    \beta}}{1 - \alpha - \beta}
\end{equation}
\begin{equation}\label{delta-morse}
    a_1 = \displaystyle \frac{1}{2 b_1}
    \left\{ \frac{b_2}{(1 - \alpha - \beta )^2 }
    \left[ 2 a_2 \left( 1 - 4 \alpha \beta \right)
    + (1 + \alpha + \beta ) \right] - b_1 \right\}
\end{equation}
Thus
\begin{equation}\label{vpm-morse}
    V_{\pm}(x) = \displaystyle
    a_1 ^2 + b_1 ^2 \exp (-2x) - b_1 \left( 2 a_1 \mp 1
    \right)\exp(-x)
    + a_2 ^2 \frac{1 - 4 \alpha \beta}{\left( 1 - \alpha - \beta \right)
    ^2}
\end{equation}
admit energies
\begin{equation}\label{morse-e}
    \begin{array}{lll}
    \varepsilon _n ^{(-)} &=&
    \displaystyle a_1 ^2 ~-~ \left( a_1-n \right) ^2  ~+~
    a_2 ^2 \frac{1 - 4 \alpha \beta }{\left( 1 - \alpha -
    \beta \right) ^2 } \ ,
    \qquad   n=0, 1, 2, \cdots < a_1  \\ \\
    \varepsilon _n ^{(+)} &=& \varepsilon _{n+1} ^{(-)}
    \end{array}
\end{equation}
The solutions of the original non Hermitian Hamiltonian $H_-$ are
given by
\begin{equation}\label{sol-morse}
    \psi _n ^{(-)} (x) \approx \displaystyle
    y^{\lambda_1 - \mu _1 - n } e^{\left(
    \frac{ \mu _2}{\delta _1} - 1 \right) \frac{y}{2} }
    L_n ^{2 \lambda_1 - 2n } (y)
\end{equation}
where $\mu _1 $ and $ \mu _2 $ are defined in equation
(\ref{mu-12}) and
\begin{equation}
    y = 2 \delta_1 e^{-x}
\end{equation}

\vspace{1cm}

\subsection{Case 4 : $g(x) = 1 \ , \ {\rm{and}} \ f(x) = x $ }

\vspace{.3cm}

\noindent These values represent the {\bf Shifted Oscillator},
denoted by the potential
\begin{equation}\label{s-osc}
    V(x) = \displaystyle \frac{a^2}{4} \left( x - \frac{2b}{a} \right)^2 -
    \frac{a}{2} \qquad , \qquad \- \infty \leq x \leq \infty
\end{equation}
with
\begin{equation}\label{w-sosc}
    W(x) = \displaystyle \frac{1}{2} a_2 x - b_2
\end{equation}
Proceeding in a similar fashion, and assuming $w(x)$ to be of the
same form as $W(x)$, with $a_2 \ , \ b_2$ replaced by $a_1 \ , \
b_1$, we obtain the following results :
\begin{equation}\label{ab-sosc}
    a_1 = \displaystyle \frac{a_2 \ \sqrt{1 - 4 \alpha \beta} }{ 1 - \alpha - \beta
    } \qquad , \qquad b_1 = \displaystyle \frac{b_2 \ \sqrt{1 - 4 \alpha \beta}
    }{ 1 - \alpha - \beta }
\end{equation}
so that
\begin{equation}\label{vpm-sosc}
    V_{\pm}(x) = \displaystyle
    \frac{1}{4} a_1 ^2 \left( x - \frac{2b_1}{a_1} \right)^2
    \pm \frac{a_1}{2} - \frac{a_2}{2 \left( 1 - \alpha - \beta \right)}
\end{equation}
with energy
\begin{equation}\label{sosc-e}
    \begin{array}{lll}
    \varepsilon _n ^{(-)} &=&
    \displaystyle a_1 n - \frac{a_2}{2 \left( 1 - \alpha - \beta \right)} \ ,
    \qquad   n=0, 1, 2, \cdots \\ \\
    \varepsilon _n ^{(+)} &=& \varepsilon _{n+1} ^{(-)}
    \end{array}
\end{equation}
Writing the solutions of $H_-$ directly
\begin{equation}\label{sol-shifted}
    \psi _n ^{(-)} (x) \approx \displaystyle
    e^{\left( \mu _1 - a_1 \right) \frac{x^2}{4}
    + \left( \mu _2 - b_1 \right) x } H_n (y)
\end{equation}
where $H_n (y)$ are the Hermite polynomials, $ y = \displaystyle
\sqrt{ \frac{a_1}{2}} \left( x - \frac{2 b_1}{a_1} \right) $ and $
\mu_1 = \mu a_2 $ , $ \mu _2 = \mu b_2 $. It can be checked that
for the solutions to behave properly in the entire interval, the
parameters should obey the condition $ \mid \alpha + \beta \mid <
1 $.

\vspace{1cm}

\section{Conclusions}

To conclude, we have developed a formalism to find an isospectral
partner Hamiltonian $H_+$ of the generalized Swanson model, viz.,
$ H_- = {\cal{A}}^{\dagger} {\cal{A}} + \alpha {\cal{A}} ^2 +
\beta {\cal{A}}^{\dagger \ 2} $. Though both the initial
Hamiltonian $H_-$ as well as its partner $H_+$ are non Hermitian,
nevertheless they have real energies for certain range of
parameter values. It is observed that $H_{\pm}$ form a pair of
pseudo super symmetric partners of a pseudo super Hamiltonian
${\cal{H}}$, and share identical energies except for the ground
state. Furthermore, the same similarity transformation operator
$\rho$ maps the pair of non Hermitian Hamiltonians $H_{\pm}$ to
their respective Hermitian counterparts $h_{\pm}$, through $
H_{\pm} = \rho ^{-1} \ h_{\pm} \ \rho $, and these Hermitian maps
form a pair of supersymmetric partners, generated by supercharges
$q, q^{\dagger}$. The pseudo super charges $Q, Q^{\sharp}$
generating the pseudo super algebra of ${\cal{H}}$ are also
related to $q, q^{\dagger}$ through the similarity transformation
: $ Q = \rho ^{-1} q \ \rho $.

Since we have introduced non Hermiticity through an imaginary
vector potential, the Hermitian maps $h_{\pm}$ obtained by
similarity transformation are Schr\"{o}dinger operators comprising
of the standard kinetic term plus a local real Hermitian
potential. It may be mentioned here that though two Hamiltonians
may be related by similarity transformations, yet they can reveal
different physical aspects of the dynamical system. In fact, for a
particular class of potentials, certain physical properties are
expected to emerge more distinctly in the non Hermitian framework.
For example, {\it exceptional points}, or branch-point
singularities of the spectrum and eigenfunctions, are associated
with non Hermitian operators \cite{EP}. However, when one goes
from the non Hermitian to the corresponding Hermitian picture, the
exceptional points are lost, and consequently the entire
information related to such phenomena. Additionally, though the
super symmetric partners $ h_{\pm}$ of a Hermitian Hamiltonian can
always be mapped to non Hermitian ones (say $H_{\pm}$) by a
similarity transformation, there is absolutely no way to determine
whether $H_{\pm}$ are isospectral or not. This is due to the fact
that to write the pseudo Hermitian partner Hamiltonian $H_+$ in
terms of the generalized annihilation and creation operators
${\cal{A}}$ and ${\cal{A}}^{\dagger}$ is still an open problem. As
a result, while $h_{\pm}$ look similar in appearance (being
expressed in terms of the creation and annihilation operators
$A^{\dagger}$ and $A$), $H_{\pm}$ are not look-alikes.
Nevertheless, we have been able to express $H_{\pm}$ in terms of
the operators $D_{\pm}$, thus proving them to be related by pseudo
super symmetry, sharing identical energies, barring the ground
state.

We have applied our formalism successfully to all the known
classes of shape-invariant models where the parameters of the
original potential and its shape-invariant partner are related
through translation. A general formula has been obtained for
generating the respective pseudo supersymmetric partner
Hamiltonians for such cases. Interestingly, the wave functions are
automatically normalized following the normalization criterion for
pseudo Hermitian systems \cite{mostafa}. We have intentionally
left out the 3-dimensional shape-invariant models falling in this
category, viz. 3-dimensional oscillator and Coulomb models, as we
have restricted this work to deal with one-dimensional systems
only. However, the radial part of these models can be studied in
this framework, with $0 \leq r \leq \infty$.

This work deals with real Hamiltonians that are nevertheless non
Hermitian. We can make a straightforward extension of our
formalism to map a complex non Hermitian Hamiltonian $H$ to a
Schr\"{o}dinger Hamiltonian which is also complex but ${\cal{PT}}$
symmetric. However, in such a case $H$ will be weakly pseudo
Hermitian \cite{weak}. Finally we would like to note that in this
work we have studied shape-invariant models with unbroken
supersymmetry. It would be interesting to study models with broken
supersymmetry, too, in this framework.

\vspace{1cm}

\section{Acknowledgment}

The authors thank the referees for their valuable comments,
without which the paper could not be written in the present form.
This work was partly supported by SERC, DST, Govt. of India,
through the Fast Track Scheme for Young Scientists
(SR/FTP/PS-07/2004), to one of the authors (AS).


\vspace{1cm}




\begin{thebibliography}{999}

\bibitem{bender} C. M. Bender \& S. Boettcher, Phys. Rev. Lett. {\bf
80} (1998) 5243. \\
C. M. Bender and S. Boettcher, J. Phys. A : Math. Gen. {\bf 31}
(1998) L273.
\bibitem{nonHerm}
M. Znojil, J. Phys. A : Math. Gen. {\bf 33} (2000) 4561. \\
G. L\'{e}vai  and M. Znojil, J. Phys. A : Math. Gen. {\bf 33} (2000) 7165. \\
P. Dorey, C. Dunning and R. Tateo, J. Phys. A : Math. Gen. {\bf
34} (2001) 5679. \\ C. M. Bender, S. Boettcher, H. F. Jones, P. N.
Meisinger and M. Simsek, Phys. Lett. A {\bf 291} (2001) 197. \\
Z. Ahmed, Phys. Lett A {\bf 282} (2001) 343, \ {\it ibid.}
Phys. Lett. A {\bf 287} (2001) 295. \\
B. Bagchi and C. Quesne, Phys. Lett. A {\bf 273} (2000) 285, \
{\it ibid.} Phys. Lett. A {\bf 300} (2002) 18.\\
C. M. Bender, D. C. Brody and H. F. Jones, Phys. Rev. Lett. {\bf
89} (2002) 270401, {\it ibid.} Phys. Rev. Lett. {\bf
92} (2004) 119902(E).\\
C. M. Bender and B. Tan, J. Phys. A : Math. Gen. {\bf 39}
(2006) 1945.\\
E. Caliceti, F. Cannata and S. Graffi, J. Phys. A : Math. Gen.
{\bf 39} (2006) 10019,  {\it and references therein}. \\
C. Quesne, J. Phys. A : Math. Theor. {\bf 40} (2007) F745-F751. \\
F. Cannata, M.V. Ioffe, D.N. Nishnianidze, Phys. Lett. A {\bf 369}
(2007) 9. \\
C. M. Bender, Rep. Prog. Phys. {\bf 70} (2007) 947.\\
Paulo E. G. Assis and A. Fring, J. Phys. A : Math. Theor. {\bf 41}
(2008) 244001. \\
Also, see link {\it
http://gemma.ujf.cas.cz/~znojil/conf/index.html} with Special
issues on {\it Conference Proceedings on pseudo-Hermitian
Hamiltonians in quantum Physics}: \\
Czech. J. Phys. {\bf 54} (2004), Czech J. Phys. {\bf 55} (2005),
J. Phys. A : Math. Gen. {\bf 39} (2006), Czech J. Phys. {\bf 56}
(2006), J. Phys. A : Math. Theor. {\bf 41} 2008.
\bibitem{susy} F. Cooper, A. Khare and U. Sukhatme, Supersymmetry
in Quantum Mechanics, {\it World Scientific}, 2001. \\
B. Bagchi, Supersymmetry in Quantum and Classical Mechanics, {\it
Chapman and Hall}, 2000.\\
H. Kalka and G. Soff, Supersymmetry, {\it Teuber}, 1997.
\bibitem{intertwine} L. Infeld and T. E. Hull, Rev. Mod. Phys. {\bf 23}
(1951) 21.
\bibitem{darboux} V. V. Fatveev and M. A. Salle, Darboux
Transformations and Solitons, {\it New York, Springer}, 1991.
\bibitem{jpa-07} A. Sinha and P. Roy, J. Phys. A : Math. Theor.
{\bf 40} (2007) 10599.
\bibitem{swanson} M. S. Swanson, J. Math. Phys. {\bf 45} (2004)
585.
\bibitem{geyer} D. P. Musumbu, H. B. Geyer
and W. D. Heiss, J. Phys. A : Math. Theor. {\bf 40} (2007) F75. \\
H. B. Geyer, I. Snyman and F. G. Scholtz, Czech. J. Phys. {\bf 54}
(2004) 1069. \\
H. F. Jones, J. Phys. A : Math. Gen. {\bf 38} (2005) 1741. \\
B. Bagchi, C. Quesne and R. Roychoudhury, J. Phys. A : Math. Gen.
{\bf 38} (2005) L647. \\
F. G. Scholtz and H. B. Geyer, Phys. Lett. B {\bf 634} (2006) 84.
\\
F. G. Scholtz and H. B. Geyer, J. Phys. A : Math. Gen. {\bf 39}
(2006) 10189. \\
H. B. Geyer, W. D. Heiss and M. Znojil, J. Phys. A : Math. Gen.
{\bf 39} (2006).
\bibitem{quesne1} C. Quesne, J. Phys. A : Math. Theor. {\bf 40}
(2007) F745.
\bibitem{quesne2} C. Quesne, J. Phys. A : Math. Theor. {\bf 41 }
(2008) 244022.
\bibitem{jones} H. F. Jones, J. Phys. A : Math. Gen. {\bf 38} (2005)
1741.
\bibitem{mostafa2} A. Mostafazadeh and A. Batal, J. Phys. A :
Math. Gen. {\bf 37 } (2004) 11645.
\bibitem{mostafa} A. Mostafazadeh, J. Math. Phys. {\bf 43}
(2002) 205, {\it ibid. } 2814, {\it ibid.} 3944.
\bibitem{mostafa-psusy} A. Mostafazadeh, Nucl. Phys. B {\bf 640}
(2002) 419.\\
M. Znojil, {\it Annihilation and creation operators in
non-Hermitian supersymmetric quantum mechanics}, arXiv:hep-th/0012002 \\
M. Znojil, {\it PT symmetry and supersymmetry},
arXiv:hep-th/0209062 \\
M. Znojil, J.Phys. A : Math. and Gen. {\bf 35} (2002) 2341. \\
M. Znojil, F. Cannata, B. Bagchi and R. Roychoudhury, Phys. Lett.
B {\bf 483} (2000) 284. \\
P. Dorey, C. Dunning and R. Tateo, J.Phys. A : Math. and Gen. {\bf
34} (2001) L391. \\
F. Kleefeld, {\it Non-Hermitian Quantum Theory and its Holomorphic
Representation: Introduction and Some Applications},
arXiv:hep-th/0408028.
\bibitem{shape-invariant} L. Gendenshtein, JETP Lett. {bf 38}
(1983) 356.
\bibitem{faria} C. F. M. Faria and A. Fring, J. Phys. A : Math.
Gen. {\bf 39} (2006) 9269.
\bibitem{EP} W.D. Heiss, {\it Exceptional Points of Non Hermitian
Operators}, arXiv : quant-ph / 0304152v1.
\bibitem{weak} L. Solombrino, {\it Weak
pseudo-Hermiticity and antilinear commutant}, arXiv:quant-ph /
0203101


\end{thebibliography}
\end{document}